\documentclass[11pt]{article}

\usepackage[margin=1in]{geometry}
\usepackage{setspace}
\usepackage{amsmath}
\usepackage{amssymb}
\usepackage{amsthm}
\usepackage{exscale}
\usepackage[mathscr]{eucal}
\usepackage{bm}
\usepackage{eqlist} 
\usepackage[final]{graphicx}
\usepackage[dvipsnames]{color}
\usepackage{verbatim}
\usepackage{natbib}
\usepackage{caption}
\usepackage{enumerate}
\usepackage{subcaption}
\usepackage{algorithm}
\usepackage{algorithmic}
\usepackage{dsfont}
\usepackage{diagbox}
\usepackage{rotating}
\usepackage{lscape}
\usepackage{url}

\usepackage{etoolbox}
\usepackage{booktabs}

\makeatletter
\patchcmd{\env@cases}{1.2}{0.6}{}{}
\makeatother

\DeclareGraphicsExtensions{.pdf, .jpg}
\DeclareMathOperator*{\argmax}{arg\,max}

\newtheorem{theorem}{Theorem}

\theoremstyle{definition}

\newcommand{\ra}[1]{\renewcommand{\arraystretch}{#1}}

\title{\Large Model-Based Clustering of Time-Evolving Networks through Temporal Exponential-Family Random Graph Models}
\author{Kevin H. Lee$^1$, Lingzhou Xue$^2$\thanks{Corresponding author.} \, and David R.~Hunter$^2$ \\ $^1$Western Michigan University and $^{2}$Pennsylvania State University }
\date{First Version: June 2016; This Version: September 2017}

\begin{document}

\maketitle

\abstract{
Dynamic networks are a general language for describing time-evolving complex systems, and discrete time
network models provide an emerging statistical technique for various applications. It is a fundamental
research question to detect the community structure in time-evolving networks. However, due to significant computational
challenges and difficulties in modeling communities of time-evolving networks, there is little progress in the current literature
to effectively find communities in time-evolving networks. In this work, we propose a novel model-based clustering framework
for time-evolving networks based on discrete time exponential-family random graph models. To choose the number of
communities, we use conditional likelihood to construct an effective model selection criterion. Furthermore, we
propose an efficient variational expectation-maximization (EM) algorithm to find approximate maximum likelihood estimates
of network parameters and mixing proportions. By using variational methods and minorization-maximization (MM) techniques,
our method has appealing scalability for large scale time-evolving networks. The power of our method is demonstrated in
simulation studies and empirical applications to international trade networks and the collaboration networks of a large
American research university.

}

\section{Introduction}
Dynamic networks are a general language for describing time-evolving complex systems, and discrete time network models
provide an emerging statistical technique to study biological, business, economic, information, and social systems in the
real world. For example, time-evolving networks shed light on understanding critical processes such as the study of
biological functions using protein-protein interaction networks \citep{han-etal-2004, taylor-etal-2009}, and also contribute to
assessing infectious disease epidemiology, dynamic brain networks and time-evolving structures of social networks
\citep{morris-kretzschmar-1995,bearman-2004,kossinets-watts-2006,park2013structural,lee2017nonparametric}.

A community can be defined as a set of nodes sharing similar connectivity patterns. In computer science and statistical
physics, many node clustering algorithms have been developed. \cite{girvan-newman-2002} propose an algorithm to
identify communities based on edge ``betweenness". They construct communities by progressively removing the edges
that connect communities most from the original network. \cite{newman-girvan-2004} proposed three different measures
of ``betweenness" and compared the results based on modularity, which measures the quality of a particular division of
a network. On the other hand, in statistics, analyzing and clustering networks is often based on statistical mixture models.
One idea of model-based clustering in networks comes from \cite{handcock-etal-2007}, who propose a latent position
cluster model that extends the latent space model of \cite{hoff-etal-2002} to take account of clustering, using the
model-based clustering ideas of \cite{fraley-raftery-2002}. In current literature, there are two very popular statistical models.
One is the stochastic block model (SBM) and the other is the exponential-family random graph model (ERGM).

Stochastic block models were first introduced by \cite{holland-etal-1983} and they focused on the case of \emph{a}
\emph{priori} specified blocks, where the memberships are known or assumed and the goal is to estimate a matrix of
edge probabilities. A statistical approach to \emph{a} \emph{posteriori} block modeling for networks was introduced
by \cite{snijders-1997} and \cite{nowicki-2001}, where the objective is to simultaneously estimate the matrix of edge
probabilities and the memberships. \cite{airoldi-etal-2008} relax the assumption of a single latent role for nodes and develop
a mixed membership stochastic block model. \cite{karrer-newman-2011} relax the assumption that a stochastic block
model treats all nodes within a community as stochastically equivalent and propose a degree-corrected stochastic block
model that can consider node covariates. Moreover, in recent years, asymptotic theory of these models has been advanced
by several pioneering papers including \cite{bickel-chen-2009}, \cite{airoldi-etal-2012}, \cite{amini-etal-2013}, and
\cite{choi-wolfe-2014}. The communities found in stochastic block models are interpreted meaningfully in many research
fields. For example, in citation and collaboration networks, such communities can be interpreted as scientific disciplines
\citep{newman-2004, ji-jin-2014}. Communities in food web networks can be interpreted as ecological subsystems
\citep{girvan-newman-2002}. Unlike the time-evolving networks considered in the current article, cross-sectional networks
are the basis of most of the stochastic block model literature cited here.

Exponential-family random graph models allow researchers to incorporate interesting features of the network into
statistical models. Moreover, researchers can specify a model capturing those features and cluster nodes based on the
specified model. Indeed, the stochastic block model is a special case of a mixture of exponential-family random graph
models. Some estimation
algorithms for exponential-family random graph models do not scale well computationally to large networks.
\cite{vu-etal-2013} propose ERGM-based clustering for large-scale cross-sectional networks that solves the scalability issue
by assuming dyadic independence conditional on the cluster memberships of nodes. In recent years, several authors
have proposed discrete time network models based on ERGM. \cite{hanneke-etal-2010} propose a temporal ERGM (TERGM)
to fit the model to a network series and \cite{krivitsky-handcock-2014} propose a separable temporal ERGM (STERGM)
that gives more flexibility in modeling time-evolving networks.

Our work is primarily motivated by detecting communities in time-evolving networks, and our results advance existing
literature by introducing a promising framework that incorporates model-based clustering while remaining
computationally scalable to large networks. This framework is based on discrete time exponential-family random graph
models and inherits the philosophy of finite mixture models, which simultaneously allows both modeling and detecting
communities in time-evolving networks, helping researchers and practitioners understand the complex structure of these
networks. Moreover, we propose a conditional likelihood Bayesian information criterion to solve the model
selection problem of determining an appropriate number of communities. We also propose an efficient variational
expectation-maximization (EM) algorithm that exhibits computational scalability for large-scale time-evolving networks by
exploiting variational methods and minorization-maximization (MM) techniques.

In Section~\ref{sec:methodology}, we present our model-based clustering method for time-evolving networks based on a finite
mixture of discrete time exponential-family random graph models. In Section~\ref{sec:selection}, we use conditional
likelihood to construct an effective model selection criterion. Section~\ref{sec:computation} designs an efficient variational
expectation-maximization algorithm to find approximate maximum likelihood estimates of network parameters and mixing
proportions. Given these estimates, we can infer membership labels and solve the problem of community detection for
time-evolving networks. The power of our method is demonstrated by simulation studies in Section~\ref{sec:simulation}
and real-world applications to international trade networks and collaboration networks in Section~\ref{sec:applications}.

\section{Methodology}
\label{sec:methodology}

\subsection{Model-based clustering of time-evolving networks through discrete time ERGMs}

In this section, we present the model-based clustering method for time-evolving networks based on a finite mixture of
discrete time exponential-family random graph models. First, we introduce some necessary notation.
We consider $n$ nodes that are
fixed over time and indexed by integers $1, \ldots, n$. Let $\mathbf{Y}_{t}=(Y_{t,ij})_{1\le i,j\le n}\in\mathcal{Y}$ represent the
network at time $t=0,1,\ldots,T$ and denote by $\mathbf{y}_{t}=(y_{t,ij})_{1\le i,j\le n}$ the corresponding observed network,
where $\mathcal{Y}$ is the set of all possible networks. Let $\boldsymbol{\theta} \in \mathbb{R}^{p}$ be a vector of $p$
network parameters of interest. Under the $k$-order Markov assumption, discrete time exponential-family random graph
models are of the form
\begin{equation}\label{tergmkorder}
\mbox{pr}_{\boldsymbol{\theta}}(\mathbf{Y}_{t} = \mathbf{y}_{t} \mid \mathbf{y}_{t-1}, \ldots, \mathbf{y}_{0}) =
\exp\{\boldsymbol{\theta}'\boldsymbol{g}(\mathbf{y}_{t}, \mathbf{y}_{t-1}, \ldots, \mathbf{y}_{t-k}) -
\psi(\boldsymbol{\theta},\mathbf{y}_{t-1}, \ldots, \mathbf{y}_{t-k})\},
\end{equation}
where $\psi(\boldsymbol{\theta}, \mathbf{y}_{t-1}, \ldots, \mathbf{y}_{t-k})$ is given by
$$
\psi(\boldsymbol{\theta}, \mathbf{y}_{t-1}, \ldots, \mathbf{y}_{t-k}) = \mbox{log}\sum_{\mathbf{y}^{*} \in
\mathcal{Y}}\mbox{exp}\left[\boldsymbol{\theta}'\boldsymbol{g}(\mathbf{y}^{*},\mathbf{y}_{t-1}, \ldots, \mathbf{y}_{t-k}) \right]
$$
and ensures that $\mbox{pr}_{\boldsymbol{\theta}}(\mathbf{Y}_{t} = \mathbf{y}_{t} \mid \mathbf{y}_{t-1}, \ldots, \mathbf{y}_{0})$
sums to $1$.
Here, $\boldsymbol{g}(\mathbf{y}_{t}, \mathbf{y}_{t-1}, \ldots, \mathbf{y}_{t-k})$
is a $p$-dimensional vector of sufficient statistics on networks $\mathbf{y}_{t}$, $\ldots$, $\mathbf{y}_{t-k}$.

We now focus on the simplest case of discrete time exponential-family random graph models under the first-order Markov
assumption and we write the one-step transition probability from $\mathbf{Y}_{t-1}$ to $\mathbf{Y}_{t}$ as
\begin{equation}\label{tergm}
\mbox{pr}_{\boldsymbol{\theta}}(\mathbf{Y}_{t} = \mathbf{y}_{t} \mid \mathbf{y}_{t-1}) =
\exp\{\boldsymbol{\theta}'\boldsymbol{g}(\mathbf{y}_{t}, \mathbf{y}_{t-1})-\psi(\boldsymbol{\theta},\mathbf{y}_{t-1})\},
\end{equation}
where $\psi(\boldsymbol{\theta}, \mathbf{y}_{t-1})$ and $\boldsymbol{g}(\mathbf{y}_{t}, \mathbf{y}_{t-1})$ are defined
as above.

\bigskip
\noindent \textbf{Remark 1}:
 Given covariates $\mathbf{x}_{t}$ and a vector $\boldsymbol{\beta} \in \mathbb{R}^{q}$ of
covariate coefficients, we can also write the transition probability from $\mathbf{Y}_{t-1}$ to $\mathbf{Y}_{t}$ with
covariates as
\begin{equation*}
\mbox{pr}_{\boldsymbol{\theta}, \boldsymbol{\beta}}(\mathbf{Y}_{t} = \mathbf{y}_{t} \mid \mathbf{y}_{t-1}, \mathbf{x}_{t}) =
\mbox{exp}\{\boldsymbol{\theta}'\boldsymbol{g}(\mathbf{y}_{t}, \mathbf{y}_{t-1}) +
\boldsymbol{\beta}'\boldsymbol{h}(\mathbf{y}_{t}, \mathbf{x}_{t})-\psi(\boldsymbol{\theta},\boldsymbol{\beta},\mathbf{y}_{t-1})\},
\end{equation*}
where $\psi(\boldsymbol{\theta},\boldsymbol{\beta}, \mathbf{y}_{t-1}) =
\log\sum_{\mathbf{y}^{*} \in \mathcal{Y}}\exp\left[\boldsymbol{\theta}'\boldsymbol{g}(\mathbf{y}^{*}, \mathbf{y}_{t-1}) +
\boldsymbol{\beta}'\boldsymbol{h}(\mathbf{y}^{*}, \mathbf{x}_{t})\right]$.
Here, $\boldsymbol{h}(\mathbf{y}_{t}, \mathbf{x}_{t})$ is a $q$-dimensional vector of statistics.

In general, for some choices of $\boldsymbol{g}(\mathbf{y}_{t}, \mathbf{y}_{t-1})$, the model in \eqref{tergm} is not tractable for
modeling large networks, since the computing time to evaluate the likelihood function directly
grows as $2^{\binom{n}{2}}$ in the case of undirected
edges. Here, we restrict our attention to scalable exponential-family models by only choosing statistics that preserve
conditional dyadic independence
wherein the distribution of $\mathbf{Y}_{t}$ given $\mathbf{Y}_{t-1}$ factors over the edge states, i.e.,
\begin{equation}\label{cdi2}
\mbox{pr}_{\boldsymbol{\theta}}(\mathbf{Y}_{t} = \mathbf{y}_{t} \mid \mathbf{y}_{t-1}) =
\prod_{i<j}^{n}\mbox{pr}_{\boldsymbol{\theta}}(Y_{t,ij} = y_{t,ij} \mid \mathbf{y}_{t-1}).
\end{equation}

Before proceeding, we introduce specific examples of statistics that preserve conditional dyadic independence and capture
interesting time-evolving network features in both TERGM and STERGM:
\begin{equation}\label{density}
g^{d}(\mathbf{y}_{t}, \mathbf{y}_{t-1}) = \sum_{i<j}^{n}y_{t,ij},
\end{equation}
\vspace{-1em}
\begin{equation}\label{stability}
g^{s}(\mathbf{y}_{t}, \mathbf{y}_{t-1}) = \sum_{i<j}^{n}\left[y_{t,ij}y_{t-1,ij} + (1-y_{t,ij})(1-y_{t-1,ij})\right].
\end{equation}
\vspace{-1em}
\begin{equation}\label{formation}
g^{f}(\mathbf{y}_{t}, \mathbf{y}_{t-1}) = \sum_{i<j}^{n}\left[y_{t,ij} - y_{t,ij}y_{t-1,ij}\right],
\end{equation}
\vspace{-1em}
\begin{equation}\label{persistence}
g^{p}(\mathbf{y}_{t}, \mathbf{y}_{t-1}) = \sum_{i<j}^{n}y_{t,ij}y_{t-1,ij}.
\end{equation}
The subscripted $i<j$ and superscripted $n$ mean that summation should be taken over all pairs $(i, j)$ with
$1\leq i < j \leq n$; the same is true for products as in equation~\eqref{cdi2}. Corresponding to the first and second statistics above are
TERGM parameters: $\theta^{d}$ relates to density, or the number of edges in the network at time $t$, while $\theta^{s}$ relates to
stability, or the number of edges maintaining their status from time $t-1$ to time $t$. Corresponding to the third and fourth
statistics above are STERGM parameters:  $\theta^{f}$ relates to formation, or the number of edges absent at time
$t-1$ but present at time $t$, while $\theta^{p}$ relates to persistence, or the number of edges existing at time $t-1$
that survive to time $t$.

Here, as in \cite{vu-etal-2013}, we assume that the probability mass function has a $K$-component mixture form as follows:
\begin{equation}\label{cdi-mixture}
\begin{split}
\mbox{pr}_{\boldsymbol{\pi}, \boldsymbol{\theta}}(\mathbf{Y}_{t} = \mathbf{y}_{t} \mid \mathbf{y}_{t-1}) =
& \sum_{\boldsymbol{z} \in \mathcal{Z}}\mbox{pr}_{\boldsymbol{\theta}}(\mathbf{Y}_{t} =
\mathbf{y}_{t} \mid \mathbf{y}_{t-1}, \boldsymbol{z})\mbox{pr}_{\boldsymbol{\pi}}(\boldsymbol{Z} = \boldsymbol{z}) \\
=& \sum_{\boldsymbol{z} \in \mathcal{Z}} \prod_{i<j}^{n}\mbox{pr}_{\boldsymbol{\theta}_{z_{i}z_{j}}}(Y_{t,ij} =
y_{t,ij} \mid \mathbf{y}_{t-1}, \boldsymbol{z})\mbox{pr}_{\boldsymbol{\pi}}(\boldsymbol{Z} = \boldsymbol{z}),
\end{split}
\end{equation}
where $\boldsymbol{Z}=(\boldsymbol{Z}_1, \ldots, \boldsymbol{Z}_n)$ denotes the membership indicators with distributions
$$
\boldsymbol{Z}_{i} \mid \pi_{1}, \ldots, \pi_{K} \overset{\text{i.i.d.}} \sim \text{Multinomial}(1; \pi_{1}, \ldots, \pi_{K})
$$
and $\mathcal{Z}$ denotes the support of $\boldsymbol{Z}$. In the mixture form \eqref{cdi-mixture}, the assumption of
conditional dyadic independence given $\boldsymbol{z}$ strikes a balance between complexity and parsimony. For now, the
number of communities $K$ is fixed and known. In Section~\ref{sec:selection}, we will discuss how to choose an optimal
number of communities $K$.

Now, we consider inference based on observing a series of networks,
$\mathbf{y}_{1}, \mathbf{y}_{2}, \ldots, \mathbf{y}_{T}$, given an initial
network $\mathbf{y}_{0}$. The log-likelihood of the observed network series is
\begin{equation}\label{loglikelihood}
\begin{split}
\ell(\boldsymbol{\pi},\boldsymbol{\theta}) =& \log \Bigg[ \prod_{t=1}^T\mbox{pr}_{\boldsymbol{\pi}, \boldsymbol{\theta}}(\mathbf{Y}_{t} =
\mathbf{y}_{t} \mid \mathbf{y}_{t-1})\Bigg]\\
=& \sum_{t=1}^T\log \left[ \sum_{\boldsymbol{z} \in \mathcal{Z}}\mbox{pr}_{\boldsymbol{\theta}}(\mathbf{Y}_{t} =
\mathbf{y}_{t} \mid \mathbf{y}_{t-1}, \boldsymbol{z})\mbox{pr}_{\boldsymbol{\pi}}(\boldsymbol{Z} = \boldsymbol{z})\right].
\end{split}
\end{equation}
Our aim is to estimate parameters $\boldsymbol{\pi}$ and $\boldsymbol{\theta}$ via maximizing the log-likelihood
$\ell(\boldsymbol{\pi},\boldsymbol{\theta})$, i.e.,
$$
(\hat{\boldsymbol{\pi}},\hat{\boldsymbol{\theta}}) =
\argmax_{(\boldsymbol{\pi}, \boldsymbol{\theta})}\ell(\boldsymbol{\pi},\boldsymbol{\theta}).
$$
Section~\ref{sec:computation} designs a novel variational EM algorithm to efficiently find the approximate maximum
likelihood estimates.  We shall see that the parameter estimates obtained by this algorithm
can provide community membership labels.

Before proceeding, we give specific examples of discrete time exponential-family random graph models with stability
parameter(s) that control the rate of evolution of a network. Stability parameters are popular in the study of time-evolving
networks; in sociology, researchers are interested in whether, say, same-gender friendships are more stable than other
friendships or whether there are differences among ethnic categories in forming lasting sexual partnerships over time
\citep{knecht-2008, krivitsky-etal-2011}.

\bigskip
\noindent \textbf{Example~1}:
When $K=1$ and $\boldsymbol{g}(\mathbf{y}_{t}, \mathbf{y}_{t-1})$ consists only of the stability statistics \eqref{stability},
the model reduces to TERGM with a stability parameter as in \cite{hanneke-etal-2010}:
\begin{equation*}
\begin{split}
&\mbox{pr}_{\boldsymbol{\theta}^s_{z_iz_j}}(Y_{t,ij} = y_{t,ij} \mid \mathbf{y}_{t-1},  \boldsymbol{z}) \\
\propto \ & \exp[(\theta_{z_i}^{s} + \theta_{z_j}^{s})(y_{t,ij}y_{t-1,ij} + (1-y_{t,ij})(1-y_{t-1,ij}))].
\end{split}
\end{equation*}

\bigskip
\noindent \textbf{Example~2}:
When $K=1$ and $\boldsymbol{g}(\mathbf{y}_{t}, \mathbf{y}_{t-1})$ consists of both formation
parameters~\eqref{formation} and persistence parameters~\eqref{persistence}, the model reduces to
STERGM with formation and persistence parameters as in \cite{krivitsky-handcock-2014}:
\begin{equation*}
\begin{split}
&\mbox{pr}_{\boldsymbol{\theta}^{f}_{z_iz_j},\boldsymbol{\theta}^{p}_{z_iz_j}}(Y_{t,ij} = y_{t,ij} \mid \mathbf{y}_{t-1},  \boldsymbol{z}) \\
\propto \ & \exp[(\theta_{z_i}^{f} + \theta_{z_j}^{f})(y_{t,ij} - y_{t,ij}y_{t-1,ij}) + (\theta_{z_i}^{p} +
\theta_{z_j}^{p})y_{t,ij}y_{t-1,ij}].
\end{split}
\end{equation*}

\subsection{Parameter Identifiability}

The unique identifiability of the parameters in a broad class of random graph mixture models has been shown by
\cite{allman-etal-2009} and \cite{allman-etal-2011}. Here we prove the generic identifiability for our proposed parameterizations.
Theorem~\ref{thm:identifiability}, whose proof is given in Appendix,
extends the identifiability result of the stochastic block model of \cite{allman-etal-2009}
and \cite{allman-etal-2011} to discrete time exponential-family random graph mixture models.
In this context, ``generically identifiable'' means uniquely identifiable except possibly on a subset of the parameter space whose
Lebesgue measure is zero.

\begin{theorem}\label{thm:identifiability}
Let $n$ be the number of nodes in a time-evolving network. The parameters $\pi_{k}$, $1\leq k \leq K$, and
the conditional probability of observing an edge $p_{kl} = \mbox{pr}_{\boldsymbol{\theta}_{kl}}(Y_{t,ij} = 1 \mid \mathbf{y}_{t-1}, \boldsymbol{z})$, $1 \leq k \leq l \leq K$ of equation \eqref{cdi-mixture} are generically identifiable, up to permutations of the subscripts $1, \ldots, K$, if
\begin{equation*}
\begin{cases}
    n^{1/2} \geq K-1 + (K+2)^{2}/4, & \text{for $K$ even};\\
    n^{1/2} \geq K-1 + (K+1)(K+3)/4, & \text{for $K$ odd}.
\end{cases}
\end{equation*}
Moreover, the network parameters $\boldsymbol{\theta}_{k} \in \mathbb{R}^{p}$, $1\leq k \leq K$ in the model (e.g., Examples~1 and~2) are
generically identifiable, up to permutations of the subscripts $1, \ldots, K$, if $p \leq \lfloor (K+1)/2 \rfloor$.
\end{theorem}

\section{Model selection}
\label{sec:selection}

In practice, the number of communities is unknown and should be chosen. \cite{handcock-etal-2007} propose a Bayesian
method of determining the number of clusters by using approximate conditional Bayes factors in a latent position cluster
model. \cite{daudin-etal-2008} also derive a Bayesian model selection criterion that is based on the integrated classification
likelihood (ICL). In this section, we use the conditional likelihood of the network series, conditioning on an estimate of the
membership vector, to construct an effective model selection criterion.

We obtain the conditional log-likelihood of the network series $\mathbf{y}_{1}, \mathbf{y}_{2}, \ldots, \mathbf{y}_{T}$, given
initial network $\mathbf{y}_{0}$ and estimated membership vector $\hat{\boldsymbol{z}}$, as
\begin{equation}
{\rm{cl}}(\boldsymbol{\theta}, \hat{\boldsymbol{z}}) = \sum_{t=1}^{T}\log [\mbox{pr}_{\boldsymbol{\theta}}(\mathbf{Y}_{t} =
\mathbf{y}_{t} \mid \mathbf{y}_{t-1}, \hat{\boldsymbol{z}})],
\end{equation}
which can be written using conditional dyadic independence \eqref{cdi2} in the form
$$
{\rm{cl}}(\boldsymbol{\theta}, \hat{\boldsymbol{z}}) = \sum_{t=1}^{T}\sum_{i<j}^{n}
\log [\mbox{pr}_{\boldsymbol{\theta}_{\hat{z}_{i}\hat{z}_{j}}}(Y_{t,ij} = y_{t,ij} \mid \mathbf{y}_{t-1}, \hat{\boldsymbol{z}})].
$$
We propose the conditional likelihood Bayesian information criterion to choose the number of communities for our
method:
\begin{equation}\label{clbic}
{\mbox{CL-BIC}}_{K} = -2{\rm{cl}}(\hat{\boldsymbol{\theta}}_{\text{mle}}, \hat{\boldsymbol{z}}) +
d_{K}(\hat{\boldsymbol{\theta}}_{\text{mle}}, \hat{\boldsymbol{z}})\log[Tn(n-1)/2],
\end{equation}
where $\hat{\boldsymbol{\theta}}_{\text{mle}}$ is the maximum likelihood estimate assuming $K$ communities and
$d_{K}({\boldsymbol{\theta}}, \hat{\boldsymbol{z}}) = \mbox{tr}(H_{K}^{-1}V_{K})$
is the model complexity, following \cite{varin-vidoni-2005}, \cite{varin2011overview} and \cite{xue2012nonconcave}, based on
$H_{K} = \mbox{E}(-\nabla_{\boldsymbol{\theta}}^{2}{{\rm{cl}}}(\boldsymbol{\theta},\hat{\boldsymbol{z}}))$ and
$V_{K}=\mbox{var}(\nabla_{\boldsymbol{\theta}}{{\rm{cl}}}(\boldsymbol{\theta},\hat{\boldsymbol{z}}))$. We choose the optimal
$K$ by minimizing the CL-BIC score.

\bigskip
\noindent \textbf{Remark 2}:
In cross-sectional networks, a similar criterion, the
composite likelihood BIC, is proposed by
\cite{saldana-etal-2014} in the stochastic block model setting where the membership vector is estimated
separately using a method such as spectral clustering.

We may derive the explicit conditional likelihood BIC for Examples~1 and~2, i.e., the TERGM and STERGM cases.
For TERGM with a stability parameter, we obtain
\begin{equation*}
\begin{split}
{\rm{cl}}(\boldsymbol{\theta}^s, \hat{\boldsymbol{z}}) =& \sum_{t=1}^{T}\sum_{i<j}^{n}
\Big[-\log(1+\exp(\theta_{\hat{z}_{i}}^{s} + \theta_{\hat{z}_{j}}^{s})) \\
& \ \ \ \ \ \ \ \ \ \ \ + (y_{t,ij}y_{t-1,ij} + (1-y_{t,ij})(1-y_{t-1,ij}))(\theta_{\hat{z}_{i}}^{s} + \theta_{\hat{z}_{j}}^{s})\Big].
\end{split}
\end{equation*}
For any given $K$ and the corresponding estimate $\hat{\boldsymbol{\theta}}_{\text{mle}}^{s}$, we derive the explicit estimate
of $V_{K}$ as
\begin{equation*}
\hat{V}_{K}(\hat{\boldsymbol{\theta}}^{s}_{\text{mle}}) = \sum_{t=1}^{T}\boldsymbol{u}
(\hat{\boldsymbol{\theta}}^{s}_{\text{mle}})\boldsymbol{u}(\hat{\boldsymbol{\theta}}^{s}_{\text{mle}})',
\end{equation*}
where $\boldsymbol{u}(\hat{\boldsymbol{\theta}}^{s}_{\text{mle}}) = (u(\hat{\theta}^{s}_{\text{mle},k}); 1 \leq k \leq K)'$ and
\begin{equation*}
\begin{split}
u(\hat{\theta}^{s}_{\text{mle},k}) =& \sum_{i<j}^{n}\Bigg[-\frac{\exp(\hat{\theta}^{s}_{\text{mle},\hat{z}_{i}}+
\hat{\theta}^{s}_{\text{mle},\hat{z}_{j}})}{1+\exp(\hat{\theta}^{s}_{\text{mle},\hat{z}_{i}}+\hat{\theta}^{s}_{\text{mle},\hat{z}_{j}})} \\
& \ \ \ \ \ \ \ + y_{t,ij}y_{t-1,ij} + (1-y_{t,ij})(1-y_{t-1,ij}) \Bigg]\left(\mathds{1}(\hat{z}_{i}=k) + \mathds{1}(\hat{z}_{j}=k)\right).
\end{split}
\end{equation*}
We also derive the explicit estimate of $H_{K}$ as
\begin{equation*}
\hat{H}_{K}(\hat{\boldsymbol{\theta}}^{s}_{\text{mle}})=
\begin{bmatrix}
T\sum_{i<j}^{n}\left[\frac{4\exp(\hat{\theta}_{\text{mle},1}^{s} + \hat{\theta}_{\text{mle},1}^{s})}{(1+
\exp(\hat{\theta}_{\text{mle},1}^{s} + \hat{\theta}_{\text{mle},1}^{s}))^{2}}\right]I_{i,j}^{1} & \ldots & T\sum_{i<j}^{n}
\left[\frac{\exp(\hat{\theta}_{\text{mle},1}^{s} + \hat{\theta}_{\text{mle},K}^{s})}{(1+\exp(\hat{\theta}_{\text{mle},1}^{s} +
\hat{\theta}_{\text{mle},K}^{s}))^{2}}\right]I_{i,j}^{1,K} \\
\vdots & \ddots & \vdots \\
T\sum_{i<j}^{n}\left[\frac{\exp(\hat{\theta}_{\text{mle},K}^{s} + \hat{\theta}_{\text{mle},1}^{s})}{(1+
\exp(\hat{\theta}_{\text{mle},K}^{s} + \hat{\theta}_{\text{mle},1}^{s}))^{2}}\right]I_{i,j}^{K,1} & \ldots & T\sum_{i<j}^{n}
\left[\frac{4\exp(\hat{\theta}_{\text{mle},K}^{s} + \hat{\theta}_{\text{mle},K}^{s})}{(1+\exp(\hat{\theta}_{\text{mle},K}^{s} +
\hat{\theta}_{\text{mle},K}^{s}))^{2}}\right]I_{i,j}^{K} \\
\end{bmatrix},
\end{equation*}
where $I_{i,j}^{k} = \mathds{1}(\hat{z}_{i}=k, \hat{z}_{j}=k)$ and $I_{i,j}^{k,l} =
\mathds{1}(\hat{z}_{i}=k, \hat{z}_{j}=l) + \mathds{1}(\hat{z}_{i}=l, \hat{z}_{j}=k)$ for $k,l=1,\ldots,K$.
We now obtain the estimate of $d_{K}$ as $\hat{d}_{K} = \mbox{tr}(\hat{H}_{K}^{-1}\hat{V}_{K})$.
Finally, for clustering time-evolving networks through TERGM with a stability parameter, we determine the optimal number of
communities from
\begin{equation}\label{cl-bic-hat}
\hat K =\arg\min_K \ \widehat{\mbox{CL-BIC}}_{K} =
\arg\min_K \  -2{\rm{cl}}(\hat{\boldsymbol{\theta}}^{s}_{\text{mle}}, \hat{\boldsymbol{z}}) +
\hat{d}_{K}(\hat{\boldsymbol{\theta}}^{s}_{\text{mle}}, \hat{\boldsymbol{z}})\log[Tn(n-1)/2],
\end{equation}
where $\hat{\boldsymbol{\theta}}^{s}_{\text{mle}}$ and $\hat{\boldsymbol{z}}$ are the estimates
of $\boldsymbol{\theta}^{s}$ and $\boldsymbol{z}$ corresponding to a given $K$. Similar details for STERGM with formation and
persistence parameters are presented in Appendix.

Here, we also introduce modified integrated classification likelihood. Again for the TERGM with a stability parameter, the
modified ICL can be written as
\begin{equation}\label{icl}
\mbox{ICL}_{K} = \sum_{t=1}^{T}\sum_{i<j}^{n}\log[\mbox{pr}_{\hat{\boldsymbol{\theta}}_{\hat{z}_{i}\hat{z}_{j}}^{s}}(Y_{t,ij} =
y_{t,ij} \mid \mathbf{y}_{t-1},  \hat{\boldsymbol{z}})] - K\log\left[Tn(n-1)/2\right],
\end{equation}
and we choose the optimal number of communities as
\begin{equation}\label{optgroup}
\hat{K} = \argmax_{K} \mbox{ICL}_{K}.
\end{equation}
We present model selection results using both conditional likelihood BIC and modified ICL in the simulation studies of
Section~\ref{sec:simulation}.

\bigskip
\noindent \textbf{Remark 3}:
Our conditional likelihood BIC can also be applied to choose the number of
communities for finite mixture of ERGMs in cross-sectional networks.
Under the assumption of conditional dyadic independence given $\boldsymbol{z}$
the conditional log-likelihood in this case can be written
as
$$
{\rm{cl}}(\boldsymbol{\theta}, \hat{\boldsymbol{z}}) = \sum_{i<j}^{n}\log [\mbox{pr}_{\boldsymbol{\theta}_{\hat{z}_{i}\hat{z}_{j}}}
(Y_{ij} = y_{ij} \mid \hat{\boldsymbol{z}})],
$$
and we choose the optimal $K$ by minimizing $\widehat{\mbox{CL-BIC}}_{K}$ as in equation~\eqref{cl-bic-hat} with $T=1$.

\section{Computation}
\label{sec:computation}

Here we present a novel variational EM algorithm to solve model-based clustering for large scale time-evolving networks. Our
algorithm is modeled on the algorithm presented by \citet{vu-etal-2013}. The algorithm combines the power of variational
methods \citep{wainwright-jordan-2008} and minorization-maximization techniques \citep{hunter-lange-2004} to effectively
handle both the computationally intractable log-likelihood function $\ell(\boldsymbol{\pi},\boldsymbol{\theta})$ and the
non-convex optimization problem of the lower bound of the log-likelihood. We introduce an auxiliary distribution
$A(\boldsymbol{z}) \equiv \mbox{pr}(\boldsymbol{Z} = \boldsymbol{z})$ to derive a tractable lower bound on the intractable log-likelihood
function. Using Jensen's inequality, the log-likelihood function may be shown to be bounded from the below as follows:
\begin{equation}\label{lowerBound}
\begin{split}
\ell(\boldsymbol{\pi},\boldsymbol{\theta}) = & \sum_{t=1}^{T}\log[\mbox{pr}_{\boldsymbol{\pi},\boldsymbol{\theta}}(\mathbf{Y}_{t} =
\mathbf{y}_{t} \mid \mathbf{y}_{t-1})] \\
= & \sum_{t=1}^{T}\log\left[\sum_{\boldsymbol{z} \in \mathcal{Z}}\frac{\mbox{pr}_{\boldsymbol{\pi},\boldsymbol{\theta}}
(\mathbf{Y}_{t}=\mathbf{y}_{t}, \boldsymbol{Z}=\boldsymbol{z} \mid
\mathbf{y}_{t-1})}{A(\boldsymbol{z})}A(\boldsymbol{z})\right] \\
\geq & \sum_{t=1}^{T}\sum_{\boldsymbol{z} \in \mathcal{Z}}\Bigg[\log \frac{\mbox{pr}_{\boldsymbol{\pi}, \boldsymbol{\theta}}
(\mathbf{Y}_{t} = \mathbf{y}_{t}, \boldsymbol{Z} = \boldsymbol{z} \mid \mathbf{y}_{t-1})}
{A(\boldsymbol{z})}\Bigg]A(\boldsymbol{z}).
\end{split}
\end{equation}

If $A(\boldsymbol{z})$ were unconstrained in the sense that we could choose from the set of all distributions with support
$\mathcal{Z}$, we would obtain the best lower bound when $A(\boldsymbol{z}) =
\mbox{pr}_{\boldsymbol{\pi}, \boldsymbol{\theta}}(\boldsymbol{Z} = \boldsymbol{z} \mid \mathbf{y}_{t}, \mathbf{y}_{t-1})$, where the
inequality becomes equality. However, this unconstrained form of $A(\boldsymbol{z})$ is intractable. We therefore constrain
$A(\boldsymbol{z})$ to a subset of tractable choices and maximize tractable lower bound to find approximate maximum
likelihood estimates.

Here, we constrain $A(\boldsymbol{z})$ to the mean-field variational family where the $\boldsymbol{Z}_{i}$ are mutually
independent,
$$
A(\boldsymbol{z}) = \prod_{i=1}^{n}\mbox{pr}_{\boldsymbol{\gamma}_{i}}(\boldsymbol{Z}_{i} = \boldsymbol{z}_{i}).
$$
We further specify $\mbox{pr}_{\boldsymbol{\gamma}_{i}}(\boldsymbol{Z}_{i} = \boldsymbol{z}_{i})$ to be
$\mbox{Multinomial}(1; \gamma_{i1}, \ldots, \gamma_{iK})$ for $i=1,\ldots,n$, where
$\boldsymbol{\Gamma} = (\boldsymbol{\gamma}_{1}, \ldots, \boldsymbol{\gamma}_{n})$ is the variational parameter. In the
estimation phase, whenever it is necessary to assign each node to a particular community, the $i$th node is assigned to the
community with the highest value among $\hat\gamma_{i1}, \ldots, \hat\gamma_{iK}$.

If we now denote the right side of Inequality~(\ref{lowerBound}) by
$\mbox{LB}(\boldsymbol{\pi}, \boldsymbol{\theta}; \boldsymbol{\Gamma})$, we may write
\begin{equation}\label{lb2}
\begin{split}
\mbox{LB}(\boldsymbol{\pi}, \boldsymbol{\theta}; \boldsymbol{\Gamma})=
& \sum_{t=1}^{T}\Bigg[\sum_{i<j}^{n}\sum_{k=1}^{K}\sum_{l=1}^{K}\gamma_{ik}\gamma_{jl}
\log[\mbox{pr}_{\boldsymbol{\theta}_{z_iz_j}}(Y_{t,ij} = y_{t,ij} \mid \mathbf{y}_{t-1}, \boldsymbol{z})] \\
& \ \ \ \ \ + \sum_{i=1}^{n}\sum_{k=1}^{K}\gamma_{ik}(\log \pi_{k} - \log \gamma_{ik})\Bigg].
\end{split}
\end{equation}
If $\boldsymbol{\pi}^{(\tau)}$, $\boldsymbol{\theta}^{(\tau)}$, and $\boldsymbol{\Gamma}^{(\tau)}$
denote the parameter estimates at the
$\tau$th iteration of our variational EM algorithm, then in principle that algorithm
consists of alternating between two steps: \\

\quad {\bf Idealized Variational E-step:}  Let $\boldsymbol{\Gamma}^{(\tau+1)} = \arg \max_{\boldsymbol\Gamma}
\mbox{LB}(\boldsymbol{\pi}^{(\tau)}, \boldsymbol{\theta}^{(\tau)}; \boldsymbol{\Gamma})$.
\smallskip
\quad {\bf Idealized Variational M-step:} Let $(\boldsymbol{\pi}^{(\tau+1)}, \boldsymbol{\theta}^{(\tau+1)}) =
\arg \max_{(\boldsymbol{\pi}, \boldsymbol{\theta})}
\mbox{LB}(\boldsymbol{\pi}, \boldsymbol{\theta}; \boldsymbol{\Gamma}^{(\tau+1)})$.

\bigskip
\noindent \textbf{Remark 4}:
If the distribution $A(\boldsymbol{z})$ were totally unconstrained, then the E-step
above would simply consist of determining the conditional distribution
$\mbox{pr}_{\boldsymbol{\pi}^{(\tau)}, \boldsymbol{\theta}^{(\tau)}}(\boldsymbol{Z}=\boldsymbol{z} \mid \mathbf{y}_{t}, \mathbf{y}_{t-1})$,
and the variational E-step and M-step would coincide with the E-step and M-step of the traditional EM algorithm
for this situation.

In our idealized variational EM algorithm, it is difficult to directly maximize the nonconcave function
$\mbox{LB}(\boldsymbol{\pi}^{(\tau)},\boldsymbol{\theta}^{(\tau)}; \boldsymbol{\Gamma})$ with respect to
$\boldsymbol{\Gamma}$. To address this challenge, we use a minorization-maximization technique to construct a
tractable minorizing function of
$\mbox{LB}(\boldsymbol{\pi}^{(\tau)},\boldsymbol{\theta}^{(\tau)}; \boldsymbol{\Gamma})$,
then maximize this minorizer.  We define
\begin{equation}\label{Qfunction}
\begin{split}
&Q(\boldsymbol{\pi}^{(\tau)}, \boldsymbol{\theta}^{(\tau)}, \boldsymbol{\Gamma}^{(\tau)}; \boldsymbol{\Gamma}) \\
= \ &\sum_{t=1}^{T}\Bigg[\sum_{i<j}^{n}\sum_{k=1}^{K}\sum_{l=1}^{K}\left(\gamma_{ik}^{2}
\frac{\gamma_{jl}^{(\tau)}}{2\gamma_{ik}^{(\tau)}} + \gamma_{jl}^{2}\frac{\gamma_{ik}^{(\tau)}}{2\gamma_{jl}^{(\tau)}}\right)
\log[\mbox{pr}_{\boldsymbol{\theta}_{z_iz_j}^{(\tau)}}(Y_{t,ij} = y_{t,ij} \mid \mathbf{y}_{t-1}, \boldsymbol{z})] \\
& \ \ \ \ \ \ + \sum_{i=1}^{n}\sum_{k=1}^{K}\gamma_{ik}\left(\log \pi_{k}^{(\tau)} - \log \gamma_{ik}^{(\tau)} -
\frac{\gamma_{ik}}{\gamma_{ik}^{(\tau)}} + 1 \right)\Bigg],
\end{split}
\end{equation}
which satisfies the defining characteristics of a minorizing function, namely
\begin{equation}\label{q1}
Q(\boldsymbol{\pi}^{(\tau)},\boldsymbol{\theta}^{(\tau)}, \boldsymbol{\Gamma}^{(\tau)}; \boldsymbol{\Gamma}) \leq
\mbox{LB}(\boldsymbol{\pi}^{(\tau)},\boldsymbol{\theta}^{(\tau)}; \boldsymbol{\Gamma})
\mbox{\ for all $\boldsymbol{\Gamma}$}
\end{equation}
and
\begin{equation}\label{q2}
Q(\boldsymbol{\pi}^{(\tau)},\boldsymbol{\theta}^{(\tau)}, \boldsymbol{\Gamma}^{(\tau)}; \boldsymbol{\Gamma}^{(\tau)}) =
\mbox{LB}(\boldsymbol{\pi}^{(\tau)},\boldsymbol{\theta}^{(\tau)}; \boldsymbol{\Gamma}^{(\tau)}).
\end{equation}
Additional details on constructing this minorizing function are presented in Appendix. Since
$Q(\boldsymbol{\pi}^{(\tau)},\boldsymbol{\theta}^{(\tau)}, \boldsymbol{\Gamma}^{(\tau)}; \boldsymbol{\Gamma})$ is concave
in $\boldsymbol{\Gamma}$ and separates into functions of the individual $\gamma_{ik}$ parameters, maximizing
$Q(\boldsymbol{\pi}^{(\tau)},\boldsymbol{\theta}^{(\tau)}, \boldsymbol{\Gamma}^{(\tau)}; \boldsymbol{\Gamma})$ is equivalent
to solving a sequence of constrained quadratic programming subproblems with respect to
$\boldsymbol{\gamma}_{1}, \ldots, \boldsymbol{\gamma}_{n}$ respectively, under constraints
$\gamma_{i1},\ldots,\gamma_{iK} \geq 0$ and $\sum_{k=1}^{K}\gamma_{ik} = 1$ for $i=1,\cdots,n$.

To maximize $\mbox{LB}(\boldsymbol{\pi}, \boldsymbol{\theta}; \boldsymbol{\Gamma}^{(\tau+1)})$ in the M-step, maximization
with respect to $\boldsymbol{\pi}$ and $\boldsymbol{\theta}$ may be accomplished separately. First, to derive the closed-form
updates for $\boldsymbol{\pi}$, we introduce a Lagrange multiplier with the constraint $\sum_{k=1}^{K}\pi_{k} = 1$. The
closed-form update for $\boldsymbol{\pi}$ is
\begin{equation}\label{optpi2}
\pi_{k}^{(\tau+1)} = n^{-1}\sum_{i=1}^{n}\gamma_{ik}^{(\tau+1)}, \quad k = 1, \ldots, K.
\end{equation}

We could obtain $\boldsymbol{\theta}^{(\tau+1)}$ using the Newton-Raphson method, though naive application of
Newton-Raphson will not guarantee an increase in \eqref{lb2} which is necessary for the ascent property of the lower
bound of the log-likelihood. Since the Hessian matrix $H(\boldsymbol{\theta}^{(\tau)})$ at the $\tau$th iteration is positive
definite, it can be easily shown that if we go in the direction
$h^{(\tau)} = -H(\boldsymbol{\theta}^{(\tau)})^{-1}\nabla \mbox{LB}(\boldsymbol{\theta}^{(\tau)}; \boldsymbol{\Gamma}^{(\tau+1)})$
of the Newton-Raphson method, we are guaranteed to go uphill initially. In our modified Newton-Raphson method we do not
find the successor point $\boldsymbol{\theta}^{(\tau+1)} = \boldsymbol{\theta}^{(\tau)} + h^{(\tau)}$ as in the standard
Newton-Raphson method. We instead take $h^{(\tau)}$ as a search direction and perform a line search \citep{Bertsimas-2009} to find
$$
\lambda^{*} = \argmax_{\lambda \in (0,1]} \mbox{LB}(\boldsymbol{\theta}^{(\tau)} + \lambda h^{(\tau)}; \boldsymbol{\Gamma}^{(\tau+1)}),
$$
then we find the successor point $\boldsymbol{\theta}^{(\tau+1)}$ by
\begin{equation}\label{NR2}
\boldsymbol{\theta}^{(\tau+1)} = \boldsymbol{\theta}^{(\tau)} - \lambda^{*}H(\boldsymbol{\theta}^{(\tau)})^{-1}\nabla
\mbox{LB}(\boldsymbol{\theta}^{(\tau)}; \boldsymbol{\Gamma}^{(\tau+1)}).
\end{equation}

Now, we summarize the details of our proposed variational EM algorithm as Algorithm 1.
\begin{algorithm}[!htpb]
\caption{{Proposed variational EM algorithm}}
\begin{itemize} \itemsep.07in
\setlength{\itemindent}{.25in}
 \item Initialize $\boldsymbol{\Gamma}^{(0)}$, $\boldsymbol{\pi}^{(0)}$, and $\boldsymbol{\theta}^{(0)}$.
 \item Iteratively solve E-step and M-step with $\tau=0, 1, 2, \ldots$ until convergence:
  \smallskip
  \begin{itemize}\itemsep.03in
\setlength{\itemindent}{.25in}
    \item \textbf{Variational E-step}: Update $\boldsymbol{\Gamma}^{(\tau+1)}$ via  maximizing $Q(\boldsymbol{\pi}^{(\tau)},
    \boldsymbol{\theta}^{(\tau)}, \boldsymbol{\Gamma}^{(\tau)}; \boldsymbol{\Gamma})$ under constraints $\gamma_{i1},\ldots,
    \gamma_{iK} \geq 0$ and $\sum_{k=1}^{K}\gamma_{ik} = 1$ for $i=1,\cdots,n$;
    \item \textbf{Variational M-step}: Compute $\pi_{k}^{(\tau+1)} = n^{-1}\sum_{i=1}^{n}\gamma_{ik}^{(\tau+1)}$ for $k=1,\ldots,K$, and
    solve $\boldsymbol{\theta}^{(\tau+1)}$ using the modified Newton-Raphson method \eqref{NR2} with the gradient and
    Hessian of \eqref{lb2}.
  \end{itemize}
\end{itemize}
\end{algorithm}

\bigskip
\noindent \textbf{Remark 5}:
The initial $\gamma_{ik}^{(0)}$ are chosen independently uniformly randomly on (0, 1), then each
$\boldsymbol{\gamma}_{i}^{(0)}$ is multiplied by a normalizing constant chosen so that
$\sum_{k=1}^{K}\gamma_{ik}^{(0)} = 1$ for every $i$. We then start with an M-step to obtain initial $\boldsymbol{\pi}^{(0)}$
and $\boldsymbol{\theta}^{(0)}$.

\bigskip
\noindent \textbf{Remark 6}:
Using standard arguments that apply to minorization-maximization, or MM algorithms
\citep{hunter-lange-2004}, we can show that our variational EM algorithm preserves
the ascent property of the lower bound of the log-likelihood, namely,
$$
\mbox{LB}(\boldsymbol{\pi}^{(\tau)}, \boldsymbol{\theta}^{(\tau)}; \boldsymbol{\Gamma}^{(\tau)}) \leq \mbox{LB}
(\boldsymbol{\pi}^{(\tau+1)}, \boldsymbol{\theta}^{(\tau+1)}; \boldsymbol{\Gamma}^{(\tau+1)}).
$$

\section{Simulation Studies}
\label{sec:simulation}

We first conduct simulation studies for a mixture of TERGMs and STERGMs. To simulate time-evolving networks from the
$K$-component mixture of TERGM with stability parameters, i.e., Example~1, we first specify network structure by choosing
randomly the categories of the nodes according to the fixed mixing proportions and by defining initial densities for each
category. Now we obtain initial network $\mathbf{y}_{0}$ by simulating all the edges between two nodes based on the
probabilities with specified density parameters and categories of the nodes. Next, we set different stability parameters for each
category and simulate all the edges in series of networks $\mathbf{y}_{1}, \ldots, \mathbf{y}_{T}$ sequentially, based on the
probabilities with specified stability parameters and given previous time point network and categories of the nodes. Similarly,
we simulate time-evolving networks from the $K$-component mixture of STERGM with formation and persistence
parameters, i.e., Example~2. For each of the four model settings listed in Table~\ref{table00}, we use 100 nodes and 10
discrete time points.

\begin{table*}[!htpb]\centering
\ra{1.3}
\caption{Model settings for TERGM with stability parameters (Model 1 and Model 2) and STERGM with formation and
persistence parameters (Model 3 and Model 4).}\label{table00}
\begin{tabular}{@{}cccccccc@{}}

& \multicolumn{3}{c}{Model 1} & & \multicolumn{3}{c}{Model 2} \\
& $\text{G}_{1}$ & & $\text{G}_{2}$ & & $\text{G}_{1}$ & $\text{G}_{2}$ & $\text{G}_{3}$ \\

Mixing proportion $\pi_k$ & 0.5 & & 0.5 & & 0.33 & 0.33 & 0.33 \\
Stability parameter $\theta^s_k$ & -0.5 & & 0.5 & & -1 & 0 & 1 \\
Initial network density parameter $\theta^d_k$ & -0.5 & & 0.5 & & -1 & 0 & 1 \\

& \multicolumn{3}{c}{Model 3} & & \multicolumn{3}{c}{Model 4} \\
& $\text{G}_{1}$ & & $\text{G}_{2}$ & & $\text{G}_{1}$ & $\text{G}_{2}$ & $\text{G}_{3}$ \\

Mixing proportion $\pi_k$ & 0.5 & & 0.5 & & 0.33 & 0.33 & 0.33 \\
Formation parameter $\theta^{f}_k$ & -1.5 & & 1.5 & & -1.5 & 0 & 1.5 \\
Persistence parameter $\theta^{p}_k$ & -1 & & 1 & & -1 & 0 & 1 \\
Initial network density parameter $\theta^d_k$ & -0.5 & & 0.5 & & -1 & 0 & 1 \\

\end{tabular}
\end{table*}

To check the performance of the algorithm at identifying the correct number of communities, we count the frequencies of
$\min\mbox{CL-BIC}$ and $\max\mbox{ICL}$ over 100 repetitions. To assess the clustering performance, we calculate the
average value of the Rand Index (RI) over the 100 repetitions \citep{rand-1971}. The measure
$\text{RI}(\boldsymbol{z}, \hat{\boldsymbol{z}})$ calculates the proportion of pairs whose estimated labels correspond to the
true labels in terms of being assigned to the same or different groups:
$$
\text{RI}(\boldsymbol{z}, \hat{\boldsymbol{z}}) = {\binom{n}{2}}^{-1}\sum_{i<j}^{n}(I\{z_{i}=z_{j}\}I\{\hat{z}_{i}=
\hat{z}_{j}\} + I\{z_{i}\neq z_{j}\}I\{\hat{z}_{i} \neq \hat{z}_{j}\}).
$$

To assess the estimation performance of the algorithm, we calculate the average $\ell^{2}$ norm loss for estimated mixing
proportions and network parameters over the 100 repetitions:
$$
\text{RSE}_{\boldsymbol{\pi}} = \left(\sum_{k=1}^{K}\left(\hat{\pi}_{k} - \pi_{k}\right)^{2}\right)^{1/2}
\quad \mbox{and} \quad
\text{RSE}_{\boldsymbol{\theta}} = \left(\sum_{k=1}^{K}\left(\hat{\theta}_{k} - \theta_{k}\right)^{2}\right)^{1/2}.
$$

First of all, we check the performance of our criterion functions
in choosing the correct number of communities. As shown in Table \ref{table11}, both
CL-BIC and modified ICL perform well.
The average Rand Index results are reported in Table \ref{table22}.

\begin{table*}[!htpb]\centering
\ra{1.3}
\caption{Frequencies of $\min\mbox{CL-BIC}$ and $\max\mbox{ICL}$ over 100 repetitions.
$K_{0}$ represents the true number of communities.}\label{table11}
\begin{tabular}{@{}cccccccccc@{}}

& \multicolumn{4}{c}{Model 1 $(K_{0}=2)$} & & \multicolumn{4}{c}{Model 2 $(K_{0}=3)$} \\
& $K=1$ & $K=2$ & $K=3$ & $K=4$ && $K=1$ & $K=2$ & $K=3$ & $K=4$ \\

$\min\mbox{CL-BIC}$ & 0 & 99 & 0 & 1 & & 0 & 0 & 97 & 3 \\
$\max\mbox{ICL}$ & 0 & 100 & 0 & 0 & & 0 & 0 & 96 & 4 \\

& \multicolumn{4}{c}{Model 3 $(K_{0}=2)$} & & \multicolumn{4}{c}{Model 4 $(K_{0}=3)$} \\
& $K=1$ & $K=2$ & $K=3$ & $K=4$ && $K=1$ & $K=2$ & $K=3$ & $K=4$ \\

$\min\mbox{CL-BIC}$ & 0 & 99 & 1 & 0 & & 0 & 3 & 93 & 4 \\
$\max\mbox{ICL}$ & 0 & 100 & 0 & 0 & & 0 & 0 & 99 & 1 \\

\end{tabular}
\end{table*}

\begin{table*}[!htpb]\centering
\ra{1.3}
\caption{Mean Rand Index values for 100 repetitions for various models and values of $K$,
with sample standard deviations in parentheses, where $K_0$ is the true number of communities.}\label{table22}
\begin{tabular}{@{}cccc@{}}

\multicolumn{4}{c}{Model 1 $(K_{0}=2)$} \\
&$K=2$ & $K=3$ & $K=4$ \\

TERGM & 1.000 (0.000) & 0.874 (0.033) & 0.773 (0.030) \\

\multicolumn{4}{c}{Model 2 $(K_{0}=3)$} \\
&$K=2$ & $K=3$ & $K=4$ \\

TERGM & 0.751 (0.044) & 0.996 (0.031) & 0.943 (0.025) \\

\multicolumn{4}{c}{Model 3 $(K_{0}=2)$} \\
&$K=2$ & $K=3$ & $K=4$ \\

STERGM & 1.000 (0.000) & 0.874 (0.033) & 0.774 (0.031) \\

\multicolumn{4}{c}{Model 4 $(K_{0}=3)$} \\
&$K=2$ & $K=3$ & $K=4$ \\

STERGM & 0.761 (0.045) & 0.998 (0.021) & 0.945 (0.018) \\

\end{tabular}
\end{table*}
Finally, Table~\ref{table33} summarizes
estimation performance of our algorithm using
$\mbox{RSE}_{\boldsymbol{\pi}}$, $\mbox{RSE}_{\boldsymbol{\theta}^{s}}$, $\mbox{RSE}_{\boldsymbol{\theta}^{f}}$,
and $\mbox{RSE}_{\boldsymbol{\theta}^{p}}$.
The results of Tables~\ref{table11} through~\ref{table33} together tell us that our algorithm performs convincingly
on this set of test datasets.

\begin{table*}[!htpb]\centering
\ra{1.3}
\caption{Average values of $\ell^{2}$ norm loss for estimated mixing proportions and network parameters over 100 repetitions
with standard deviations shown in parentheses. $K_{0}$ represents the true number of communities.}\label{table33}
\begin{tabular}{@{}ccccccc@{}}

\multicolumn{3}{c}{Model 1 $(K_{0}=2)$} & & \multicolumn{3}{c}{Model 2 $(K_{0}=3)$} \\
$\mbox{RSE}_{\boldsymbol{\pi}}$ & & $\mbox{RSE}_{\boldsymbol{\theta}^{s}}$ && $\mbox{RSE}_{\boldsymbol{\pi}}$ & &
$\mbox{RSE}_{\boldsymbol{\theta}^{s}}$ \\

0.056 (0.043) & & 0.013 (0.008) && 0.073 (0.043) & & 0.038 (0.098) \\

\multicolumn{3}{c}{Model 3 $(K_{0}=2)$} & & \multicolumn{3}{c}{Model 4 $(K_{0}=3)$} \\
$\mbox{RSE}_{\boldsymbol{\pi}}$ & $\mbox{RSE}_{\boldsymbol{\theta}^{f}}$ & $\mbox{RSE}_{\boldsymbol{\theta}^{p}}$ &&
$\mbox{RSE}_{\boldsymbol{\pi}}$ & $\mbox{RSE}_{\boldsymbol{\theta}^{f}}$ & $\mbox{RSE}_{\boldsymbol{\theta}^{p}}$ \\

0.057 (0.043) & 0.030 (0.020) & 0.023 (0.015) && 0.072 (0.040) & 0.045 (0.094) & 0.039 (0.071) \\

\end{tabular}
\end{table*}

To check which model selection criterion is more robust in choosing correct number of communities when the time-evolving
networks are not simulated from the true model, we conduct another simulation study. This time we use the `simulate.stergm'
function in the `tergm' package \citep{krivitsky2016} in R \citep{r2016} to simulate $K$ time-evolving networks
\citep{krivitsky-handcock-2014} and combine the $K$ time-evolving networks into single time-evolving networks where each
simulated time-evolving networks representing $K$ different communities. First, we specify each network structure by
choosing randomly the categories of the nodes according to the fixed mixing proportions and by defining the network
densities. Next, we set different mean relational durations, which represent different degrees of stability, and simulate each
time-evolving network to have the average network density we defined over the time points. Finally, we combine the $K$
time-evolving networks into single time-evolving networks by adding a fixed number of edges between randomly chosen pairs of
individuals in different communities. For each of the two model settings listed in Table~\ref{table44}, we use 100 nodes, 10
discrete time points, and 10 edges added between randomly chosen pairs of nodes in different communities.

\begin{table*}[!htpb]\centering
\ra{1.3}
\caption{Model settings for generating time-evolving networks using `simulate.stergm' function.}\label{table44}
\begin{tabular}{@{}cccccccc@{}}

& \multicolumn{3}{c}{Model 5} & & \multicolumn{3}{c}{Model 6} \\
& $\text{G}_{1}$ & & $\text{G}_{2}$ & & $\text{G}_{1}$ & $\text{G}_{2}$ & $\text{G}_{3}$ \\

Mixing proportion & 0.4 & & 0.6 & & 0.3 & 0.4 & 0.3 \\
Mean relational duration & 5 & & 2.5 & & 7.5 & 5 & 2.5 \\
Average network density & 0.15 & & 0.1 & & 0.1 & 0.25 & 0.3 \\

\end{tabular}
\end{table*}

As shown in Table \ref{table55}, in the new simulation setting where the time-evolving networks are not simulated from the true
model, CL-BIC still performs well in choosing the correct number of communities through both TERGM with a stability
parameter and STERGM with formation and persistence parameters.  However, as shown in Table \ref{table66}, modified ICL
fails to choose the correct number of communities. The results of Tables~\ref{table55} and~\ref{table66} together tell us that
the performance of our proposed CL-BIC in choosing the correct number of communities is more robust than modified ICL
when the model assumptions are violated, at least in the particular testing regime we implemented.

\begin{table*}[!htpb]\centering
\ra{1.3}
\caption{Frequencies of $\min\mbox{CL-BIC}$ over 100 repetitions of fitting both TERGM with stability parameter and STERGM
with formation and persistence parameters. The true number of communities is $K_{0}$.}\label{table55}
\begin{tabular}{@{}cccccccccc@{}}

& \multicolumn{4}{c}{Model 5 $(K_{0}=2)$} & & \multicolumn{4}{c}{Model 6 $(K_{0}=3)$} \\
& $K=1$ & $K=2$ & $K=3$ & $K=4$ && $K=1$ & $K=2$ & $K=3$ & $K=4$ \\

TERGM & 0 & 87 & 12 & 1 & & 0 & 1 & 96 & 3 \\
STERGM & 0 & 93 & 2 & 5 & & 0 & 8 & 91 & 1 \\

\end{tabular}
\end{table*}

\begin{table*}[!htpb]\centering
\ra{1.3}
\caption{Frequencies of $\max\mbox{ICL}$ over 100 repetitions for both TERGM with stability parameter and STERGM with
formation and persistence parameters. The true number of communities is $K_{0}$.}\label{table66}
\begin{tabular}{@{}cccccccccc@{}}

& \multicolumn{4}{c}{Model 5 $(K_{0}=2)$} & & \multicolumn{4}{c}{Model 6 $(K_{0}=3)$} \\
& $K=1$ & $K=2$ & $K=3$ & $K=4$ && $K=1$ & $K=2$ & $K=3$ & $K=4$ \\

TERGM & 0 & 49 & 23 & 28 & & 0 & 0 & 65 & 35 \\
STERGM & 0 & 53 & 32 & 15 & & 0 & 0 & 59 & 41 \\

\end{tabular}
\end{table*}

The average Rand Index results are also reported in Table \ref{table77}. In all models, both TERGM with a stability parameter
and STERGM with formation and persistence parameters achieve a high average Rand Index for the correct number of
mixtures. Moreover, we see a fairly high average Rand Index with the selected (via minimum CL-BIC) number of communities
$\hat{K}$. The results of Table \ref{table55} and \ref{table77} together tell us that our algorithm based on CL-BIC performs
convincingly in choosing the correct number of communities and assigning nodes to communities even when the time-evolving
networks are not generated from the true model.

\begin{table*}[!htpb]\centering
\ra{1.3}
\caption{Comparison of clustering performance using average Rand Index for both TERGM and STERGM models with
standard deviations shown in parentheses. The true number of communities is $K_{0}$.}\label{table77}
\begin{tabular}{@{}ccccc@{}}

\multicolumn{5}{c}{Model 5 $(K_{0}=2)$} \\
&$K=2$ & $K=3$ & $K=4$ & $K=\hat{K}$ \\

TERGM & 0.976 (0.032) & 0.797 (0.045) & 0.716 (0.036) & 0.948 (0.080) \\
STERGM & 0.979 (0.025) & 0.798 (0.054) & 0.727 (0.041) & 0.966 (0.056) \\

\multicolumn{5}{c}{Model 6 $(K_{0}=3)$} \\
&$K=2$ & $K=3$ & $K=4$ & $K=\hat{K}$\\

TERGM & 0.753 (0.054) & 0.976 (0.033) & 0.935 (0.023) & 0.975 (0.037) \\
STERGM & 0.756 (0.055) & 0.972 (0.036) & 0.931 (0.024) & 0.961 (0.052) \\

\end{tabular}
\end{table*}

\section{Applications to Real-World Time-Evolving Networks}
\label{sec:applications}

Here, we apply our proposed model-based clustering methods to detect communities in two time-evolving network datasets:
International trade networks of $58$ countries from 1981 to 2000, and collaboration networks of $151$ researchers at a large
American research university from 2004 to 2013. In particular, we are interested in analyzing the
rate of evolution of these time-evolving networks as in \cite{knecht-2008},
\cite{snijders-etal-2010}, and \cite{krivitsky-handcock-2014}. Before proceeding, we introduce metrics to measure the
instability of edges in the estimated communities $\text{G}_1,\ldots,  \text{G}_{\hat K}$. For $k,l=1,\ldots,\hat K$ and
$t=1,\ldots, T$, we first define
\begin{itemize}
\item the ``$1\rightarrow0$'' instability of edges between $\text{G}_k$ and $\text{G}_l$ at the time $t$:
$$\mathcal{S}_{1\rightarrow0}^{kl}(t) = \frac{\sum\limits_{i \in \text{G}_{k}, j \in \text{G}_{l}}y_{t-1,ij}(1-y_{t,ij})}
{\sum\limits_{i \in \text{G}_{k}, j \in \text{G}_{l}}y_{t-1,ij}y_{t,ij}};$$
\item the ``$0\rightarrow1$'' instability of edges between $\text{G}_k$ and $\text{G}_l$ at the time $t$:
$$\mathcal{S}_{0\rightarrow1}^{kl}(t) = \frac{\sum\limits_{i \in \text{G}_{k}, j \in \text{G}_{l}}(1-y_{t-1,ij})y_{t,ij}}
{\sum\limits_{i \in \text{G}_{k}, j \in \text{G}_{l}}(1-y_{t-1,ij})(1-y_{t,ij})};$$
\item the total instability of edges between $\text{G}_k$ and $\text{G}_l$ at the time $t$:
$$\mathcal{S}_{\text{tot}}^{kl}(t) = \frac{\sum\limits_{i \in \text{G}_{k}, j \in \text{G}_{l}}\bigl[y_{t-1,ij}(1-y_{t,ij})+
(1-y_{t-1,ij})y_{t,ij}\bigr]}{\sum\limits_{i \in \text{G}_{k}, j \in \text{G}_{l}}\bigl[y_{t-1,ij}y_{t,ij} + (1-y_{t-1,ij})(1-y_{t,ij})\bigr]}.$$
\end{itemize}
The three instability statistics defined above evaluate the within-group instability when $k=l$ and the between-group instability
when $k\neq l$. Next, we define $\mathcal{AS}^{kl}_{1\rightarrow0}$, $\mathcal{AS}^{kl}_{0\rightarrow1}$ and
$\mathcal{AS}^{kl}_{tot}$ as the averages over all $t$ of $\mathcal{S}_{1\rightarrow0}^{kl}(t)$,
$\mathcal{S}_{0\rightarrow1}^{kl}(t)$, and $\mathcal{S}_{tot}^{kl}(t)$, respectively. Here, a larger value of
$\mathcal{AS}^{kl}_{1\rightarrow0}$ indicates that the network is more likely to dissolve ties, a larger value of
$\mathcal{AS}^{kl}_{0\rightarrow1}$ implies that the network is more likely to form ties, and a larger value of
$\mathcal{AS}^{kl}_{tot}$ implies that the network is less stable overall.

\subsection{International trade networks}

We first consider finding communities for the yearly international trade networks of $n=58$ countries studied by
\cite{ward2007}. We follow \cite{westveld-hoff-2011} and \cite{saldana-etal-2014} to define networks
$\mathbf{y}_{1981}, \ldots, \mathbf{y}_{2000}$ as follows: for any $t=1981,\ldots,2000$, $y_{t,ij} = 1$ if the bilateral trade
between country $i$ and country $j$ in year $t$ exceeds the median bilateral trade in year $t$, and $y_{t,ij} = 0$ otherwise.
By definition, this setup results in networks in which the edge density is roughly one half.
We employ model-based clustering using a TERGM with a stability parameter, i.e., Example~1.

\begin{figure}[!htpb]
    \centering
        \includegraphics[scale=0.6]{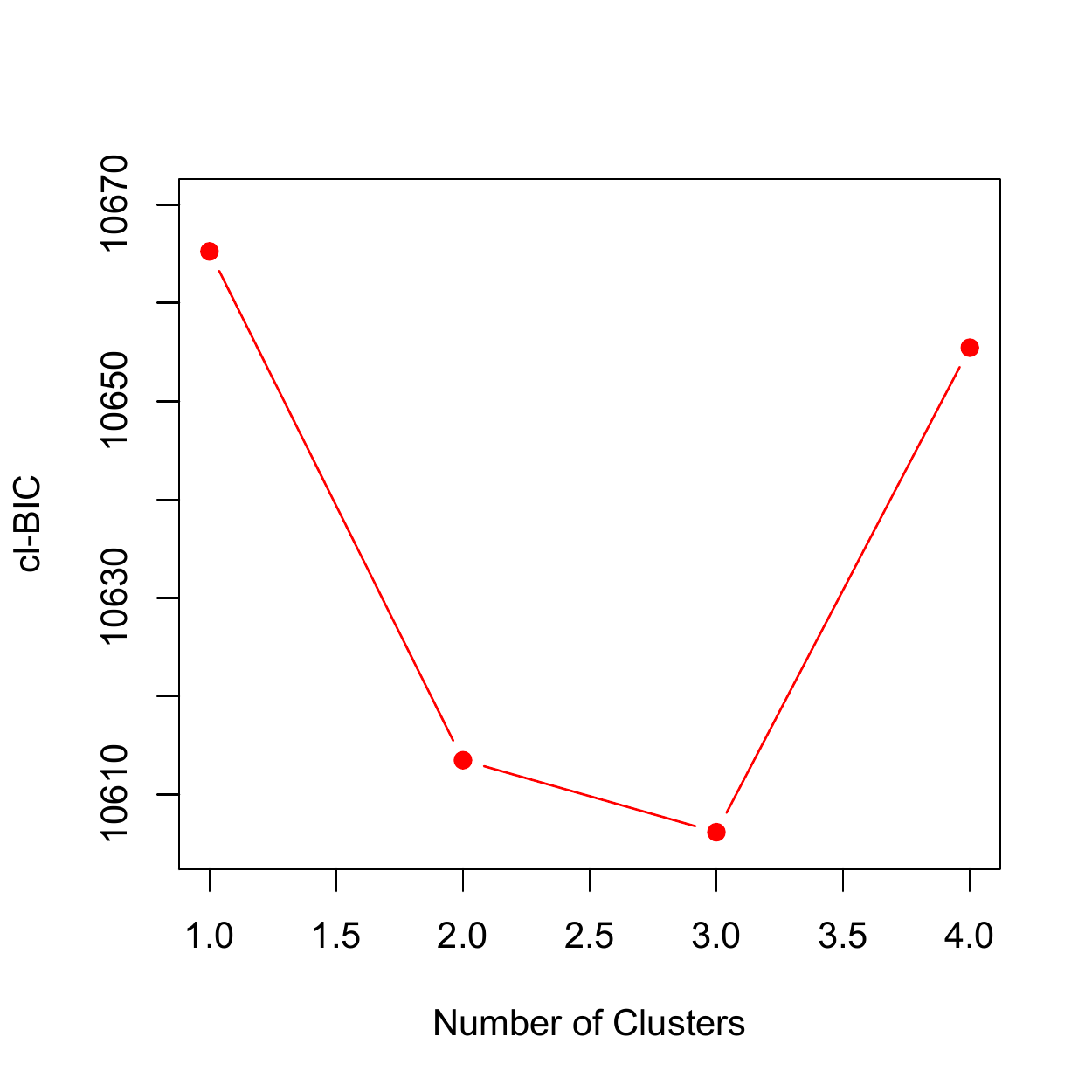}
    \caption{CL-BIC for model-based clustering through TERGM with a stability parameter in international trade networks.}
    \label{fig:clBIC1}
\end{figure}

First, we use our proposed CL-BIC to determine the number of communities. As shown in Figure \ref{fig:clBIC1},
we will use a value of $\hat K=3$.

We summarize the characteristics of the three estimated communities, including some basic network statistics and parameter
estimates, in Table~\ref{pe1}. We also calculate the within-group and between-group instability measures. As shown in
Table~\ref{tcl1}, community $\text{G}_{3}$ has the smallest total instability $\mathcal{AS}^{33}_{\text{tot}}$, which implies that
those countries in  $\text{G}_{3}$ consistently maintain their trading countries. In addition, the ``$1\rightarrow0$'' and
``$0\rightarrow1$'' instability measures show that countries in community $\text{G}_{2}$ change their trading countries more
actively than countries in $\text{G}_{1}$ and $\text{G}_{3}$. To sum up, communities $\text{G}_{1}$, $\text{G}_{2}$, and
$\text{G}_{3}$ correspond to the medium stability, low stability, and high stability groups, respectively.

\begin{table*}[!htpb]\centering
\ra{1.3}
{\small
\caption{Summary of basic network statistics, parameter estimates, and average memberships.}\label{pe1}
\begin{tabular}{@{}rccc@{}}
& $\text{G}_{1}$  & $\text{G}_{2}$ & $\text{G}_{3}$  \\
Total $\#$ of nodes & 24 & 21 & 13 \\
Average $\#$ of edges per node & 17.11 & 34.28 & 42.42 \\
Average $\#$ of triangles per node & 150.17 & 443.95 & 613.15 \\
Estimated mixing proportion $\hat{\pi}$ & 0.3920 & 0.3677 & 0.2403 \\
Estimated stability parameter $\hat{\theta}^{s}$ & 1.6323 & 1.3168 & 2.0712 \\
Average membership of $\hat{\gamma}_{\cdot1}$ & 0.8147 & 0.1158 & 0.0579 \\
Average membership of $\hat{\gamma}_{\cdot2}$ & 0.1060 & 0.8262 & 0.1101 \\
Average membership of $\hat{\gamma}_{\cdot3}$ & 0.0794 & 0.0579 & 0.8320 \\

\end{tabular}
}
\end{table*}

\begin{table}[!htpb]\centering
\ra{1.3}
{\small
\caption{Summary of within-group and between-group instability statistics for the proposed model-based clustering community
assignments for the international trade network dataset. $\mathcal{AS}^{kl}_{1\rightarrow0}$,
$\mathcal{AS}^{kl}_{0\rightarrow1}$, and $\mathcal{AS}^{kl}_{tot}$ measure the average over all $t$ of
$\mathcal{S}_{1\rightarrow0}^{kl}(t)$, $\mathcal{S}_{0\rightarrow1}^{kl}(t)$, and $\mathcal{S}_{tot}^{kl}(t)$, respectively, with
standard deviations shown in parentheses.}\label{tcl1}
\begin{tabular}{@{}ccccccccc@{}}
$\mathcal{AS}^{11}_{1\rightarrow0}$ & $\mathcal{AS}^{11}_{0\rightarrow1}$ & $\mathcal{AS}^{11}_{\text{tot}}$ &
$\mathcal{AS}^{22}_{1\rightarrow0}$ & $\mathcal{AS}^{22}_{0\rightarrow1}$ & $\mathcal{AS}^{22}_{\text{tot}}$ &
$\mathcal{AS}^{33}_{1\rightarrow0}$ & $\mathcal{AS}^{33}_{0\rightarrow1}$ & $\mathcal{AS}^{33}_{\text{tot}}$ \\
 0.088 & 0.014 & 0.023 & 0.048 & 0.248 & 0.082 & 0.002 & 0.014 & 0.004\\
 (0.094) & (0.009) & (0.013) & (0.041) & (0.132) & (0.047) & (0.006) & (0.034) & (0.006) \\

$\mathcal{AS}^{12}_{1\rightarrow0}$ & $\mathcal{AS}^{12}_{0\rightarrow1}$ & $\mathcal{AS}^{12}_{\text{tot}}$ &
$\mathcal{AS}^{13}_{1\rightarrow0}$ & $\mathcal{AS}^{13}_{0\rightarrow1}$ & $\mathcal{AS}^{13}_{\text{tot}}$ &
$\mathcal{AS}^{23}_{1\rightarrow0}$ & $\mathcal{AS}^{23}_{0\rightarrow1}$ & $\mathcal{AS}^{23}_{\text{tot}}$ \\
0.098 & 0.042 & 0.058 & 0.041 & 0.048 & 0.043 & 0.011 & 0.049 & 0.016 \\
 (0.030) & (0.014) & (0.015) & (0.021) & (0.026) & (0.008) & (0.017) & (0.088) & (0.019)\\

\end{tabular}
}
\end{table}

\begin{figure}[!htpb]
\centering
\includegraphics[width=1\linewidth]{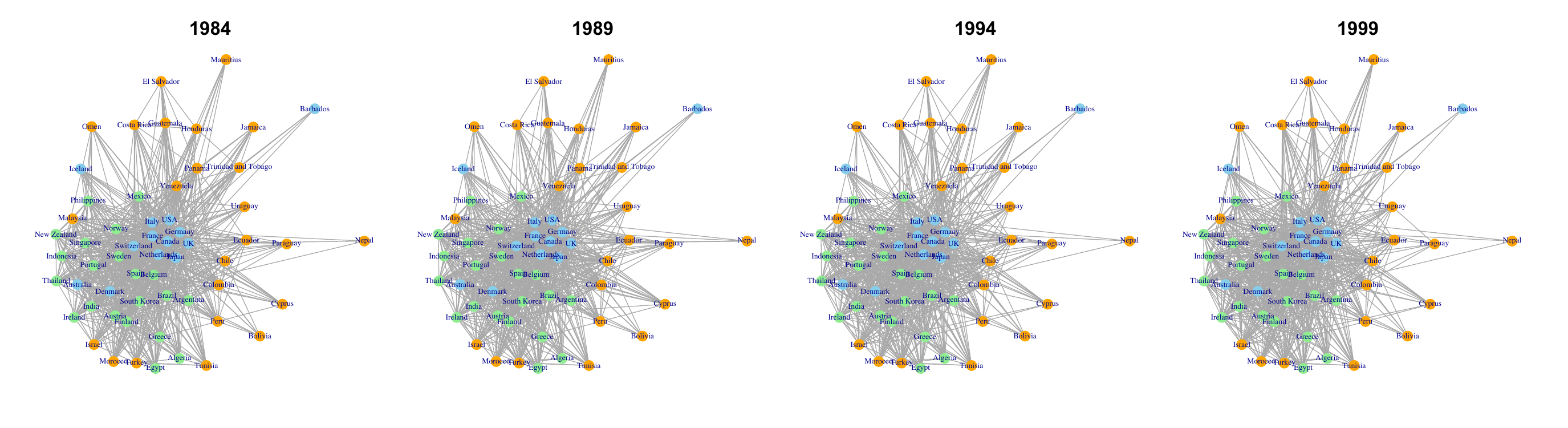}
\caption{
International trade networks with estimated communities in four different years. Nodes assigned to
$\text{G}_{1}$, $\text{G}_{2}$, and $\text{G}_{3}$ are colored orange, green, and blue, respectively.
}
\label{fig:tradeplots}
\end{figure}

In Figure~\ref{fig:tradeplots}, we plot the international trade networks with estimated communities to illustrate our model-based
clustering result. To illustrate how networks change over time for countries in each of the three communities, in
Appendix we isolate one representative country from each community: Israel from
$\text{G}_{1}$, Thailand from $\text{G}_{2}$, and the United States from $\text{G}_{3}$.

\subsection{Collaboration networks}

\begin{figure}[!htpb]
    \centering
        \includegraphics[scale=0.6]{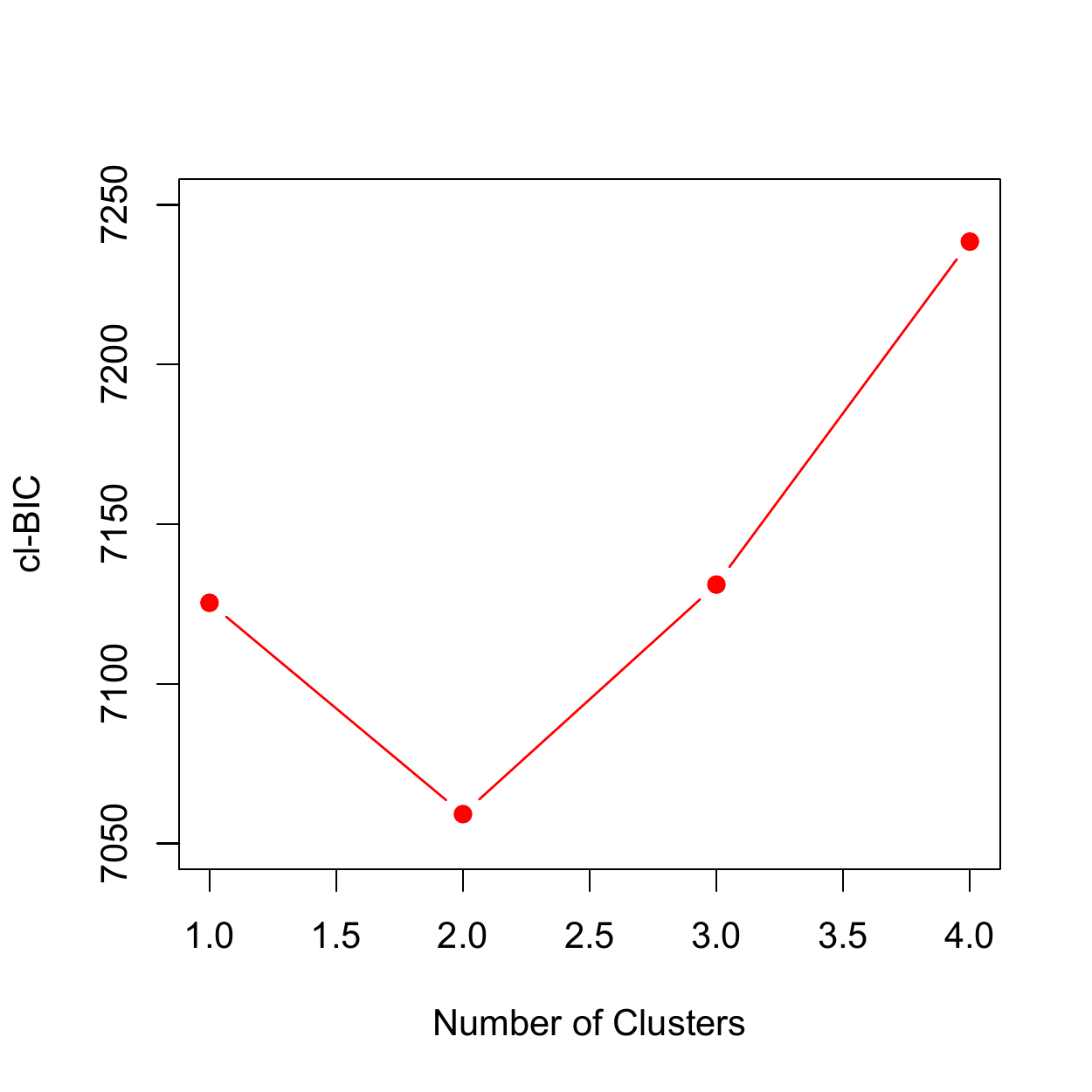}
    \caption{CL-BIC for model-based clustering using STERGMs with formation and persistence parameters in
    collaboration networks.}\label{fig:clBIC2}
\end{figure}

We next find communities for the yearly collaboration networks at a large research university from 2004 to 2013. There are
$n=151$ researchers from various academic units in this dataset. We define networks
$\mathbf{y}_{2004}, \ldots, \mathbf{y}_{2013}$ as follows: for any $t=2004,\ldots,2013$, $y_{t,ij} = 1$ if researcher $i$ and
researcher $j$ have an active research grant together during year $t$, and $y_{t,ij} = 0$ otherwise. We employ mixtures of
both TERGMs with  stability parameters, i.e., Example~1, and STERGMs with formation and persistence parameters,
i.e., Example~2. As shown in Figure \ref{fig:clBIC2}, our proposed CL-BIC indicates that the optimal number of
communities is $\hat K=2$. We shall identify two researcher communities based on different degrees of stability.

We obtain the same estimated communities from clustering through TERGM with stability parameter or STERGM with
formation and persistence parameters. Table~\ref{pe3} summarizes their basic network statistics and parameter estimates,
while Table~\ref{tcl3} displays the within-group and between-group instability measures. As these tables show,
$\text{G}_{1}$ has higher ``$1\rightarrow0$'', ``$0\rightarrow1$'', and total instability than $\text{G}_{2}$. Thus, the researchers
in $\text{G}_{1}$ tend to have fewer stable collaborations and work with more collaborators than those in $\text{G}_{2}$.

Compared to TERGMs with stability parameters, STERGMs
with formation and persistence parameters provide more detailed insights about time-evolving networks. Based on the
parameter estimates in Table \ref{pe3}, in each community, the stability is more explained by the persistence parameter than
the formation parameter. In view of this fact, we further calculate the mean relational duration using the persistence parameter
estimates for each estimated community. We obtain the mean relational durations of 2.18 years for $\text{G}_{1}$ and 2.85
years for $\text{G}_{2}$.

\begin{table*}[!htpb]\centering
\ra{1.3}
\caption{Summary of basic network statistics, parameter estimates, and average memberships.}\label{pe3}
\begin{tabular}{@{}rcc@{}}
& $\text{G}_{1}$  & $\text{G}_{2}$ \\
Total $\#$ of nodes & 34 & 117 \\
Average $\#$ of edges per node & 2.92 & 1.90 \\
Average $\#$ of triangles per node & 0.8088 & 0.3957 \\
&\multicolumn{2}{c}{\footnotesize{STERGM}}\\
Estimated mixing proportions $\hat{\pi}$ & 0.2464 & 0.7536  \\
Estimated formation parameter $\hat{\theta}^{f}$ & -2.2677 & -2.9634  \\
Estimated persistence parameter $\hat{\theta}^{p}$ & 0.1647 & 0.6156  \\
Average membership of $\hat{\gamma}_{\cdot1}$ & 0.7706 & 0.0941 \\
Average membership of $\hat{\gamma}_{\cdot2}$ & 0.2294 & 0.9059 \\
&\multicolumn{2}{c}{\footnotesize{TERGM}}\\
Estimated mixing proportions $\hat{\pi}$ & 0.3910 & 0.6090  \\
Estimated stability parameter $\hat{\theta}^{s}$ & 2.0635 & 2.7038  \\
Average membership of $\hat{\gamma}_{\cdot1}$ & 0.8462 & 0.1071 \\
Average membership of $\hat{\gamma}_{\cdot2}$ & 0.1538 & 0.8929 \\

\end{tabular}
\end{table*}

\begin{table*}[!htpb]\centering
\ra{1.3}
{\small
\caption{Summary of within-group and between-group instability statistics for the proposed model-based clustering community
assignments for the collaboration network dataset. $\mathcal{AS}^{kl}_{1\rightarrow0}$, $\mathcal{AS}^{kl}_{0\rightarrow1}$,
and $\mathcal{AS}^{kl}_{tot}$ measure the average over all $t$ of $\mathcal{S}_{1\rightarrow0}^{kl}(t)$,
$\mathcal{S}_{0\rightarrow1}^{kl}(t)$, and $\mathcal{S}_{tot}^{kl}(t)$, respectively, with standard deviations shown in
parentheses.
}\label{tcl3}
\begin{tabular}{@{}ccccccccc@{}}
 $\mathcal{AS}^{11}_{1\rightarrow0}$ & $\mathcal{AS}^{11}_{0\rightarrow1}$ & $\mathcal{AS}^{11}_{\text{tot}}$ &
 $\mathcal{AS}^{22}_{1\rightarrow0}$ & $\mathcal{AS}^{22}_{0\rightarrow1}$ & $\mathcal{AS}^{22}_{\text{tot}}$ &
 $\mathcal{AS}^{12}_{1\rightarrow0}$ & $\mathcal{AS}^{12}_{0\rightarrow1}$ & $\mathcal{AS}^{12}_{\text{tot}}$ \\
\multicolumn{9}{c}{\footnotesize{STERGM}}\\
 0.672 & 0.014 & 0.029 & 0.272 & 0.003 & 0.005 & 0.540 & 0.005 & 0.010 \\
 (0.359) & (0.006) & (0.007) & (0.059) & (0.001) & (0.001) & (0.189) & (0.001) & (0.002) \\
\multicolumn{9}{c}{\footnotesize{TERGM}}\\
 0.585 & 0.011 & 0.023 & 0.250 & 0.003 & 0.006 & 0.444 & 0.003 & 0.006 \\
 (0.211) & (0.004) & (0.005) & (0.066) & (0.001) & (0.001) & (0.076) & (0.001) & (0.001) \\

\end{tabular}
}
\end{table*}

\begin{figure}[!htpb]
\centering
\includegraphics[width=1\linewidth]{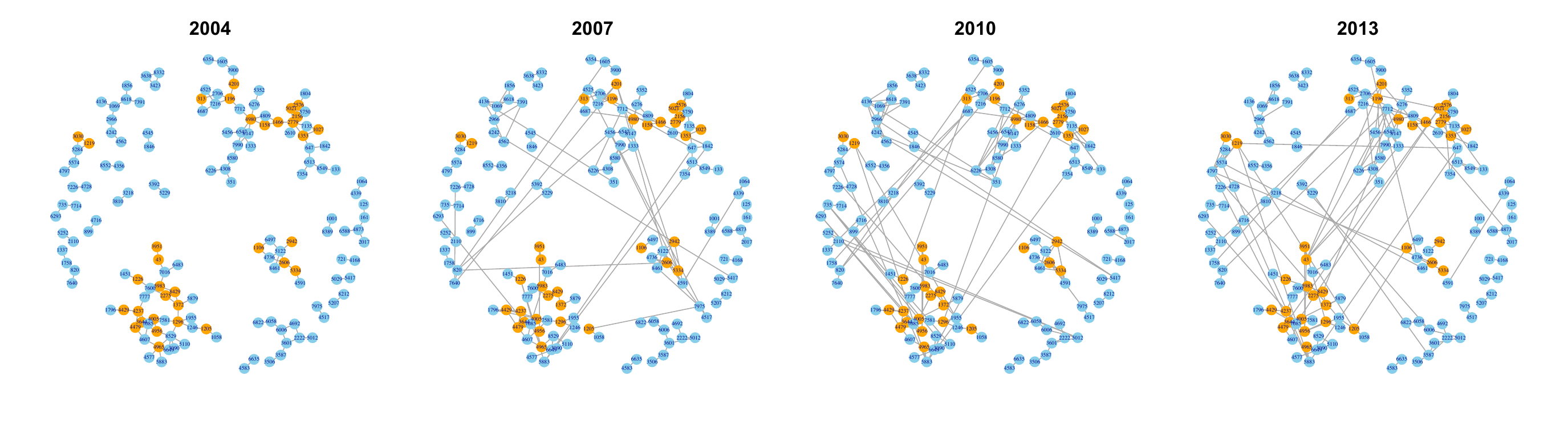}
\caption{Collaboration networks with estimated communities in four different years. Nodes assigned to $\text{G}_{1}$ and
$\text{G}_{2}$ are colored orange and blue, respectively.}
\label{fig:collaborationplots}
\end{figure}

Figure \ref{fig:collaborationplots} presents the collaboration networks with estimated communities represented by orange for
$\text{G}_{1}$ and blue for $\text{G}_{2}$. We also plot the networks of several representative individual researchers in Appendix,
anonymized by the assignment of four-digit identification numbers.

\bibliographystyle{agsm}

\end{document}